\documentclass{article}

\usepackage{arxiv}

\usepackage[utf8]{inputenc}            % allow utf-8 input
\usepackage[T1]{fontenc}               % use 8-bit T1 fonts
\usepackage[hidelinks]{hyperref}       % hyperlinks
\usepackage{url}                       % simple URL typesetting
\usepackage{booktabs}                  % professional-quality tables
\usepackage{amsfonts}                  % blackboard math symbols
\usepackage{nicefrac}                  % compact symbols for 1/2, etc.
\usepackage{microtype}                 % microtypography
\usepackage{lipsum}		               % Can be removed after putting your 
\usepackage{tabularx}
\usepackage{rotating}

\usepackage{graphicx}
\usepackage[numbers,sort&compress]{natbib}
\usepackage{doi}
\usepackage{caption}

\title{Design Perspective on Materials Experience: A CiteSpace-Based Bibliometric and Visual Analysis of Interdisciplinary Research}

\author{ 
    \includegraphics[scale=0.06]{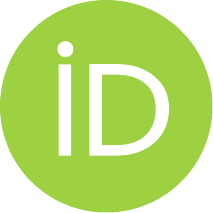}\hspace{1mm}Yuxin Zhang
    \thanks{Yuxin Zhang, PhD in Design, focuses on industrial design, material experience, affective engineering, and ergonomics. For further academic details, please refer to his ORCID profile: https://orcid.org/0000-0001-8100-3649} \\
    Academy of Arts \& Design \\
    Tsinghua University \\
    Haidian  District, Beijing, 100084, China \\
    \texttt{zhangyuxin@mail.tsinghua.edu.cn} \\
    \And
	Fan Zhang \\
	Academy of Arts \& Design \\
	Tsinghua University \\
	Haidian  District, Beijing, 100084, China \\
	\texttt{zhang-f24@mails.tsinghua.edu.cn} \\
}

\renewcommand{\shorttitle}{MATERIAL EXPERIENCE: A BIBLIOMETRIC STUDY}

\hypersetup{
pdftitle={A template for the arxiv style},
pdfsubject={q-bio.NC, q-bio.QM},
pdfauthor={David S.~Hippocampus, Elias D.~Striatum},
pdfkeywords={First keyword, Second keyword, More},
}

\begin{document}
\maketitle

\begin{abstract}
	Based on a bibliometric analysis of literature from 2005 to 2024, this study reveals that material experience is undergoing a profound transformation characterized by evolving material definitions, methodological advances, and increasing interdisciplinary integration. Material types now extend beyond traditional substances to encompass virtual and biological media, underscoring a growing emphasis on perception and interaction. Methodologically, the field has transitioned from subjective descriptions to data-driven, quantifiable models focused on objective sensory analysis and multisensory integration to enhance immersion. Key drivers, including human-machine perception convergence, material-driven interface interactions, and the embedding of intelligent interactive functions, propel the discipline toward an experience-centered paradigm reflecting a deep convergence of design, science, and technology. At the national/regional level, the United States, China, Japan, Germany, and the Netherlands lead in contributions, while France, the United Kingdom, and Romania demonstrate significant interdisciplinary progress. At the institutional level, Delft University of Technology, Justus Liebig University Giessen, and the Centre National de la Recherche Scientifique show significant advantages. In particular, the Material-Driven Design theory has established a foundational impact on the discipline, while, regarding general research trends, scholars from the United States, the Netherlands, and Germany maintain the highest academic visibility. Overall, material experience research is at a critical juncture, its future development will depend on progress in material innovation, technological integration, and perceptual quantification, as well as the establishment of socio-cultural values, all of which must be effectively unified through design to address complex evolving needs.
\end{abstract}

% keywords can be removed

\section{Introduction}

In academic research, literature reviews are essential for understanding research problems, identifying gaps, and advancing knowledge \cite{carrera-riveraHowtoConductSystematic2022, snyderLiteratureReviewResearch2019}. Each study or perspective can inspire new ideas, and even minor details may serve as catalysts for future research \cite{zouEvaluationTrendFashion2022}. Therefore, systematic thinking and continuous integration are crucial for generating new knowledge \cite{kunischReviewResearchScientific2023}. However, with the rapid growth of publications, traditional review methods have become increasingly cumbersome \cite{byrneImprovingPeerReview2016}. To address this challenge, quantitative methods have emerged, with bibliometric analysis serving as a mature and rigorous tool. It transforms textual features into analyzable numerical data, revealing research hotspots, structural relationships, and dynamic trends \cite{hooverQuantitativeAnalysisLiterary2013}. This method not only allows for the systematic examination of large-scale scientific data but also uncovers the development trajectories of fields and emerging research frontiers \cite{donthuHowConductBibliometric2021}.

As an emerging and complex research domain, material experience emphasizes a human-centered approach, encouraging exploration from multiple perspectives \cite{wilkesMaterialsLibraryCollections2018}. While significant advancements have been made across various subfields, comprehensive and systematic studies are still lacking \cite{giaccardiFoundationsMaterialsExperience2015a}. Current research is characterized by thematic diversity, methodological variation, and interdisciplinary integration, spanning design, psychology, human-computer interaction, and materials science \cite{ahmadsayutiEmotionalResponsesPerceptions2021}. In this interdisciplinary landscape, design serves as the core of material experience, playing a pivotal role, particularly in integrating material functionality with aesthetics. Its central mission lies in closely aligning the physical properties of materials with the psychological needs of users, a role that is especially prominent in product design and user experience design \cite{alanExploringMaterialPotentials2025}. Furthermore, design education focuses on cultivating future designers' deep understanding and application of material experience \cite{zhouEngagingMaterialEducation2021}. By systematically analyzing material experience, it helps design students master the use of design thinking to select appropriate materials, thereby enhancing the overall user experience \cite{renardCultivatingDesignThinking2014}. Additionally, it guides students to recognize the future direction of design, particularly in the context of rapid advancements in new materials and technologies. This enables students to anticipate future trends and adequately prepare for forthcoming design practices, thus promoting the interdisciplinary development and innovation within the design field \cite{peiDesignEducation402023}.

With the continuous expansion of research topics, the overall structure of the field has exhibited fragmented characteristics, lacking a systematic knowledge map and a clear developmental trajectory \cite{baiConstructionKnowledgeGraph2025}. This fragmentation impedes researchers from fully grasping the evolution of the field and limits both the cumulative development of theory and effective interdisciplinary integration. To facilitate the integrated development of theory and practice, bibliometric methods, by constructing keyword co-occurrence networks, co-citation maps, and burst detection visualizations, can map knowledge structures and reveal overlooked research gaps \cite{wangVisualizingKnowledgeStructure2022}. Among bibliometric tools, CiteSpace is widely adopted for its powerful visualization and data processing capabilities, making it particularly suitable for analyzing developmental trends in emerging interdisciplinary fields \cite{xuResearchDigitalManagement2024}. Through multidimensional data mining and dynamic visualizations, CiteSpace effectively reveals the intrinsic connections and knowledge flow paths among publications, providing researchers with both macro- and micro-level insights into complex academic networks \cite{chenCiteSpaceIIDetecting2006}.

This study aims to use CiteSpace to conduct a visual analysis of literature on material experience, revealing the field’s developmental trajectory, structural characteristics, and future trends. The objective is to help researchers in the design domain gain a comprehensive understanding of the knowledge system and academic landscape of material experience, and to promote the transition of material experience research from a traditional foundation toward interdisciplinary development.

\section{Method}
\subsection{Theoretical Background}

The development of bibliometrics has always revolved around a core concept—citation. Citing previous research not only establishes crucial connections among scholars, journals, and institutions—thereby constructing an empirical network suitable for quantitative analysis—but also creates an evolutionary chain in the temporal dimension between publications \cite{mejiaExploringTopicsBibliometric2021}.

Co-citation analysis, by identifying documents that are cited together, helps to uncover thematic clusters and the intellectual foundations of a research field \cite{smallCocitationScientificLiterature1973}. Co-word analysis, based on the co-occurrence frequency of terms within literature, reveals the semantic associations and structural features among research topics, offering insights into their internal relationships and evolutionary trajectories \cite{klarinHowConductBibliometric2024}. Meanwhile, co-authorship analysis focuses on collaborative relationships among scholars, highlighting the trend of cooperation in scientific research, and reveals the network expansion mechanisms and their key role in knowledge production \cite{newmanStructureScientificCollaboration2001}.

In addition, bibliometrics indicators such as degree centrality, betweenness centrality, and eigenvector centrality further quantify the relative importance of various research entities (e.g., authors, institutions, and countries/regions) within academic networks, thereby providing deeper and more multidimensional perspectives for bibliometric analysis \cite{borgattiGraphtheoreticPerspectiveCentrality2006}.

\subsection{Bibliometric Method and Visualization}

In bibliometric research, two commonly used analysis methods are performance analysis and science mapping. Performance analysis focuses on evaluating the development of a research field from the perspective of scientific output, quantifying the research performance of different projects or entities using a series of bibliometric indicators based on publications and citations \cite{gutierrez-salcedoBibliometricProceduresAnalyzing2017}. In contrast, science mapping constructs networks among authors, papers, concepts, and citations to reveal the knowledge structure, scholarly connections, and evolutionary trends of a discipline, providing a more systematic perspective for understanding the development trajectory and emerging frontiers of a research field \cite{coboScienceMappingSoftware2011}.

In the performance analysis of this study, descriptive statistics were conducted on the annual number of publications, publication titles, and research areas. In addition, we summarized the classification distributions of the target literature across four attributes: “Web of Science Categories”, “Citation Topics Meso”, “Citation Topics Micro” and “Sustainable Development Goals”. Data visualizations were created using Python (version 3.12.7). In the science mapping analysis, the data were analyzed using the advanced version of CiteSpace 6.4.R1. To ensure data completeness, the “Full Record and References” option was selected during the data download process \cite{vandagriffWeKnowWhat2023}, and duplicate records were checked using the software’s built-in deduplication function. Based on this, author collaboration networks, institutional collaboration networks, and country/region collaboration networks were constructed and analyzed to explore collaboration patterns and their development trends at different levels. The node selection criterion was set to the g-index, with a scaling factor of k = 25. To effectively simplify the network structure and highlight core relationships, commonly used pruning methods such as the Pathfinder algorithm, Pruning Sliced Networks, and Pruning the Merged Networks were applied during the analysis \cite{chenSearchingIntellectualTurning2004}. Subsequently, clustering analysis was performed using the Louvain method, with cluster labels extracted through the log-likelihood ratio (LLR) method. The quality of clustering was evaluated using the modularity, which ranges from -1 to 1, with values closer to 1 indicating a more distinct community structure, and scores above 0.7 generally considered to represent strong clustering \cite{newmanFindingEvaluatingCommunity2004, blondelFastUnfoldingCommunities2008}. In addition, keyword bursts were identified and analyzed to reveal the rapid changes and evolutionary trends of research hotspots.

During the visualization process, the size of a node reflects the frequency of occurrence or citation count of the element it represents (such as Authors, Institutions, Countries/Regions); larger nodes indicate greater importance within the field. The thickness of the links between nodes represents the strength of the relationship between them, such as co-citation frequency or co-occurrence frequency, with thicker links indicating stronger connections. The colors of nodes and links are often used to represent temporal information or cluster categories, helping researchers identify the temporal evolution of research hotspots and distinguish between different research topics, thereby enhancing the readability and interpretability of the maps \cite{chenCiteSpaceIIDetecting2006}.

\subsection{Data Process}

This study conducted a literature search on July 12, 2025, using the Web of Science Core Collection database, accessed via Tsinghua University Library. As a comprehensive platform featuring high-impact and authoritative academic journals worldwide, Web of Science offers robust support for retrieving high-quality, multidisciplinary literature \cite{birkleWebScienceData2020}. An advanced search strategy was employed, utilizing a Topic Search approach that includes fields such as title, abstract, author keywords, and Keyword Plus. To ensure strong relevance between the search results and the research topic, this study strategically applied wildcard characters (e.g., “*”) and Boolean operators (e.g., OR and AND), thereby flexibly adjusting keyword combinations to expand or narrow the search scope and improve the accuracy and breadth of the literature retrieval \cite{zupicBibliometricMethodsManagement2015}. The specific search query is as follows:

TS = (((“Material* Experience” OR “Material* Perception” OR “Material* Aesthetics”) OR (“Material*” AND (“Impression Evaluation” OR “Sensory Evaluation” OR “Kansei Engineering” OR “Affective Engineering” OR “HCI”)))).

This query was designed to comprehensively capture literature involving keywords such as “Material Experience”, “Material Perception” and “Material Aesthetics”, while also incorporating research topics related to “Impression Evaluation”, “Sensory Evaluation”, “Kansei Engineering”, “Affective Engineering” and “HCI.” The search was limited to English-language journal articles and conference proceedings published between January 1, 2005, and December 31, 2024, covering nearly two decades of research in the relevant field. The time slicing was set to one year. Although accurate search terms were used in the initial retrieval, concerns remain about whether the dataset fully represents the relevant literature \cite{ozturkHowDesignBibliometric2024}. To minimize such uncertainties, the research team conducted two rounds of manual screening on the initially retrieved 2,831 documents and ultimately selected 575 publications that closely aligned with the research objectives as the foundation for subsequent analysis.

\begin{figure}[h]
\centering
\includegraphics[width=0.8\linewidth]{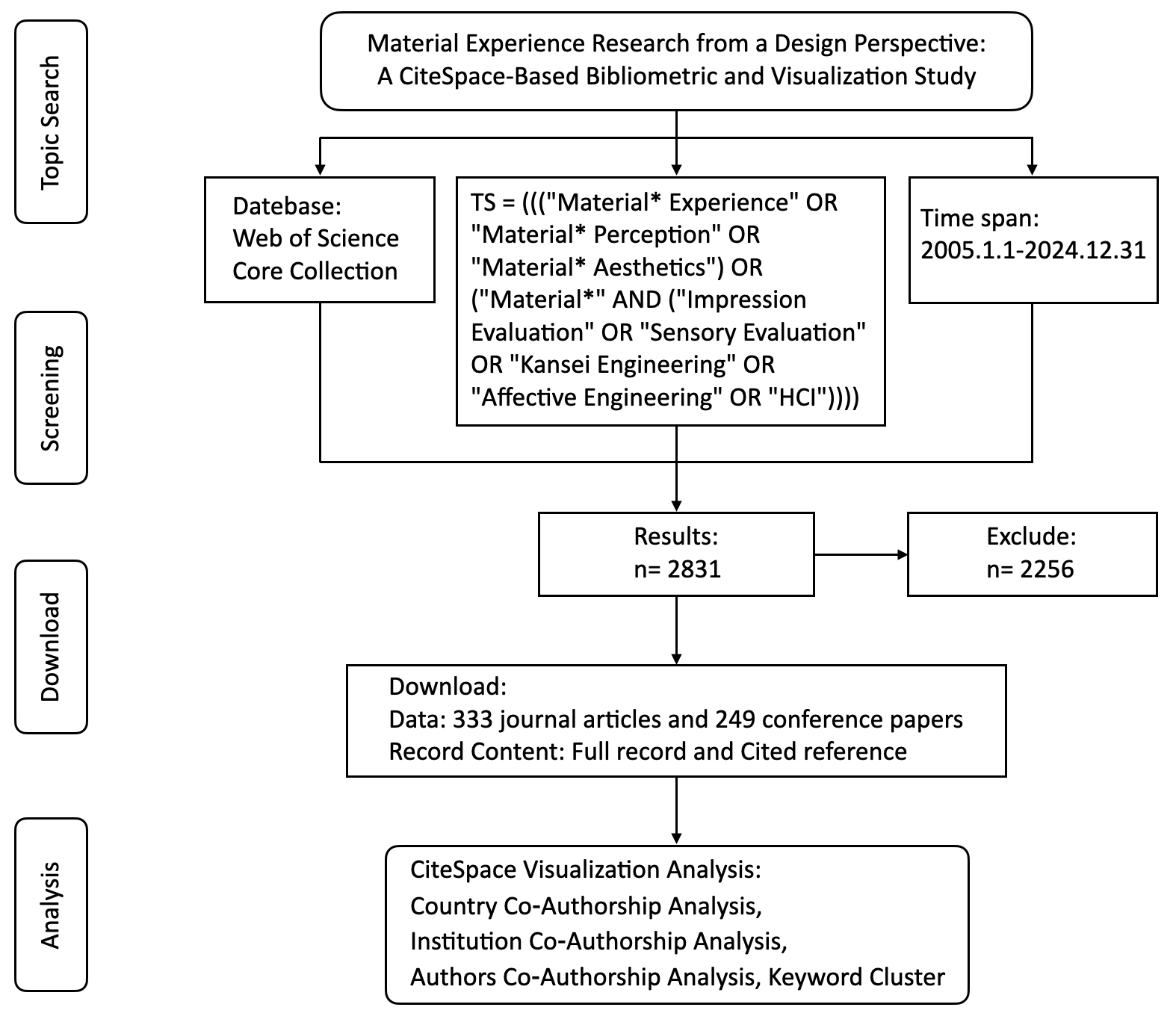}
\caption{Flowchart of the CiteSpace-Based Bibliometric and Visualization Workflow for Mapping Trends and Patterns in Material Experience Research}
\label{figure-1}
\end{figure}

Based on the selected 575 publications, this study conducted both performance analysis and science mapping. The performance analysis focused on descriptive statistics of the annual number of publications, publication titles, research areas, and the classification distribution of the target literature across four attributes: “Web of Science Categories”, “Citation Topics Meso”, “Citation Topics Micro” and “Sustainable Development Goals”. The science mapping concentrated on four key dimensions: Authors, Institutions, Countries/Regions and Keywords, with corresponding co-occurrence networks constructed for each. Subsequently, cluster analysis and timeline were performed on the keywords, along with burst term detection. The process of the science mapping analysis is shown in Figure~\ref{figure-1}.

\section{Results and Discussion}
\subsection{Publications in the last 20 years}

\begin{figure}[h]
\centering
\includegraphics[width=\linewidth]{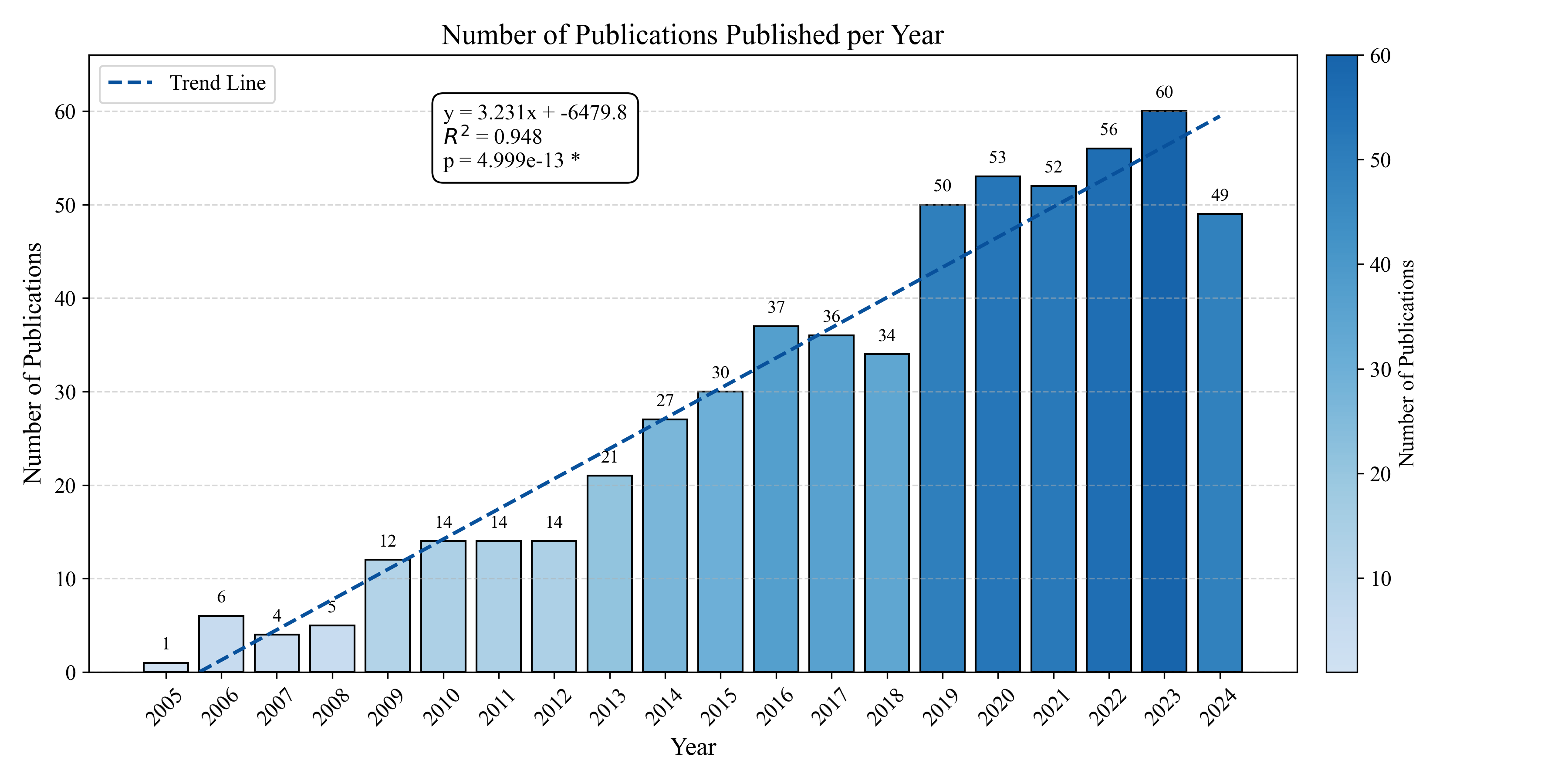}
\caption{The trend in annual publication number related to material experience over the past 20 years indicates that research in this field has shown a steady upward momentum over time}
\label{figure-2}
\end{figure}

Figure~\ref{figure-2} illustrates the annual number of journal articles and conference papers on material experience published over the past two decades. In total, 333 journal articles and 249 conference papers were identified. While journal articles are numerically predominant, conference papers play an indispensable role in fostering academic exchange and advancing research in this field. Based on data from 2005 to 2024, a linear regression analysis was performed, with $x$ representing the year and $y$ denoting the number of publications. The resulting regression model was $y = 3.231x - 6479.8$, with a regression coefficient of $\beta = 3.231$ ($p < 0.001$) and a coefficient of determination $R^2 = 0.948$. This indicates that the independent variable $x$ accounts for 94.8\% of the variance in the dependent variable $y$, revealing a significant and sustained upward trend in publication output. Such a steady growth trajectory not only demonstrates that “material experience” continues to garner scholarly attention, but also underscores its growing potential value for both academic inquiry and practical application.

It is noteworthy that publications surged in 2009 and then remained relatively stable. From 2013 onward, publication counts rose steadily. Another peak occurred in 2019, followed by a generally fluctuating upward trend. By 2024, the number of publications had slightly declined. We consider that the 2009 surge was primarily driven by the emergence of methodologies such as Kansei engineering and multisensory design in the early 21st century \cite{levyKanseiEngineeringEmancipation2013, nagamachiKanseiEngineeringPowerful2002, spenceSensesPlaceArchitectural2020}, whereas the 2019 growth was largely propelled by policy initiatives and funding \cite{nationalsciencefoundationNSF19526Materials2019}.

\begin{figure}[htbp]
\centering
\includegraphics[width=0.95\linewidth]{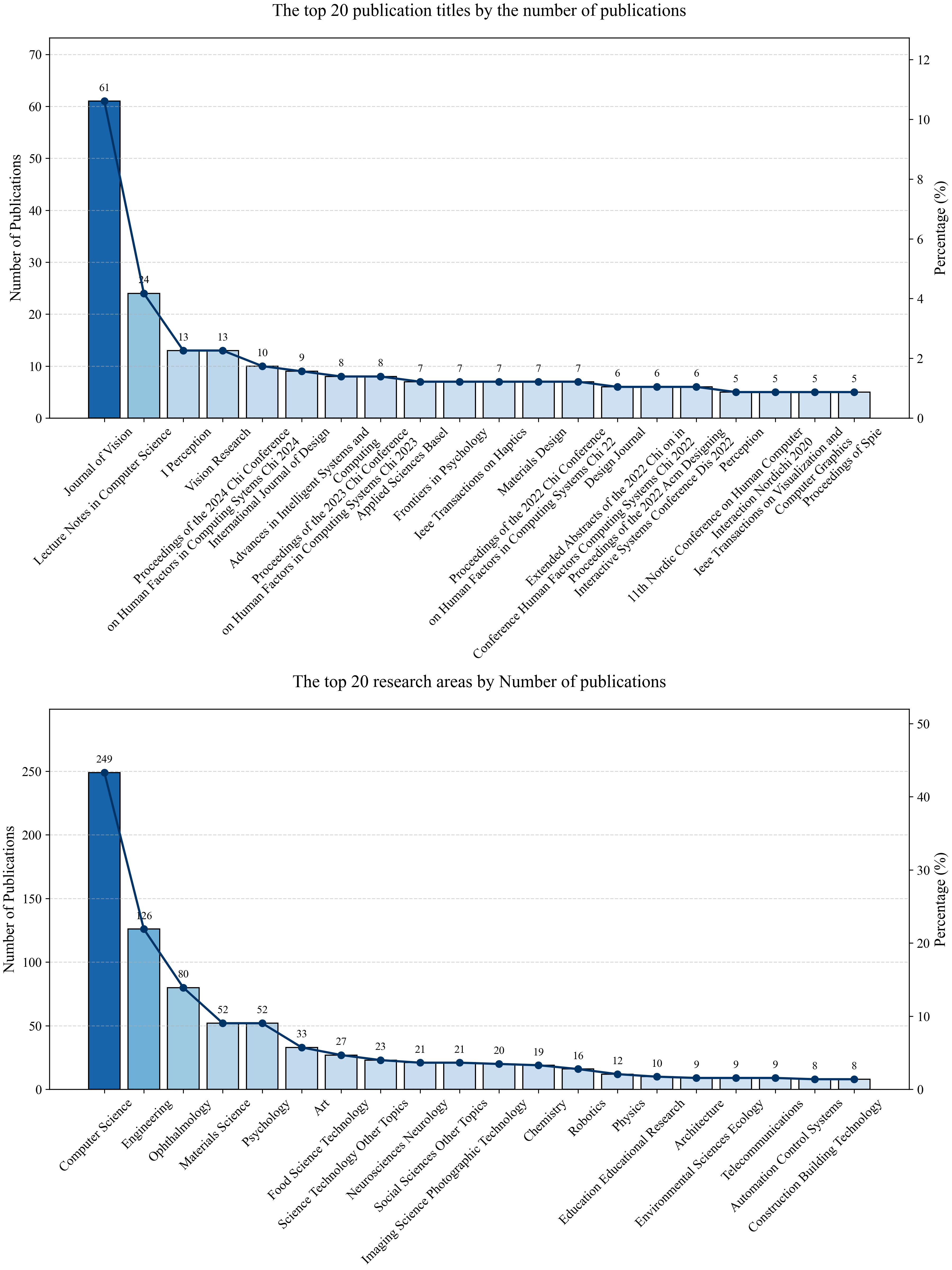}
\caption{Top 20 publication titles by the number of publications on material experience research, and top 20 research areas in Web of Science. In the figure, the left vertical axis represents the number of publications, while the right vertical axis indicates the percentage of the total relevant literature}
\label{figure-3}
\end{figure}

Figure~\ref{figure-3} (top) presents the top 20 journals and conference proceedings in terms of publication volume within material experience research, while Figure~\ref{figure-3} (bottom) shows the top 20 research areas in this field according to Web of Science. Among all publications, the Journal of Vision leads with 61 relevant articles, underscoring the central role of visual perception in material experience research as an authoritative journal in the field of vision science. Ranked second is Lecture Notes in Computer Science, a prominent conference proceedings series in computer science, with 24 related papers. Close behind are i-Perception and Vision Research, each contributing 13 articles. The former emphasizes open-access studies in perceptual science, while the latter is a classic journal in the domain of vision science.

It is worth noting that, after aggregating the CHI—Conference on Human Factors in Computing Systems—proceedings from 2022 to 2024 (including the 2022 Extended Abstracts), a total of 31 papers related to material experience were identified. This highlights CHI’s pivotal role as one of the most influential international conferences in human-computer interaction and its significant contribution to this research direction.

In addition, the International Journal of Design ranks sixth with nine articles and is one of the most influential international journals in the field of design research. Advances in Intelligent Systems and Computing ranks seventh with eight related publications, focusing on cutting-edge research in computational intelligence, artificial intelligence, and machine learning. Frontiers in Psychology reflects the sustained interest of the psychology community in the topic of material experience, while IEEE Transactions on Haptics specializes in research related to haptic interaction, virtual reality, and robotic interaction. Applied Sciences—Basel and Materials \& Design, originating from the fields of engineering and applied sciences, and materials engineering and product design respectively, each published seven relevant articles, highlighting the interdisciplinary integration of materials science and design practice.

Other notable publications on the list include: The Design Journal (6 publications), Proceedings of the 2022 ACM Designing Interactive Systems Conference (DIS 2022) (6 publications), Proceedings of the 11th Nordic Conference on Human-Computer Interaction (NordiCHI 2020) (5 publications), IEEE Transactions on Visualization and Computer Graphics (5 publications), Perception (5 publications), and Proceedings of SPIE (5 publications).

In terms of research disciplines, computer science, engineering, and ophthalmology rank as the top three, with 249, 126, and 80 publications respectively. Materials science and psychology each account for 52 publications, while the arts contribute 33. These results indicate that research on material experience is gradually moving beyond its traditional reliance on materials science, shifting toward an interdisciplinary exploration that deeply integrates art and science. Current research increasingly focuses on combining material engineering with innovative design concepts to create diverse material forms encompassing both physical and virtual dimensions. These materials stimulate physiological perception through their sensory and physical properties, thereby eliciting subjective experiences \cite{becerraUseMaterialsScience2025, luoExpectationsHumansHave2024}. At the same time, researchers are placing growing emphasis on the computability of such experiences, aiming to translate abstract emotional perceptions into explicit mathematical models \cite{bertheauxEffectMaterialProperties2024, kowalczukComputationalApproachesModeling2016, zhangSimulatingEmotionsIntegrated2024}. This interdisciplinary convergence is embedded throughout the research process. Overall, these findings further confirm the significant academic value and vast potential for development that material experience holds as a highly interdisciplinary field.

\begin{figure}[htbp]
\centering
\includegraphics[width=\linewidth]{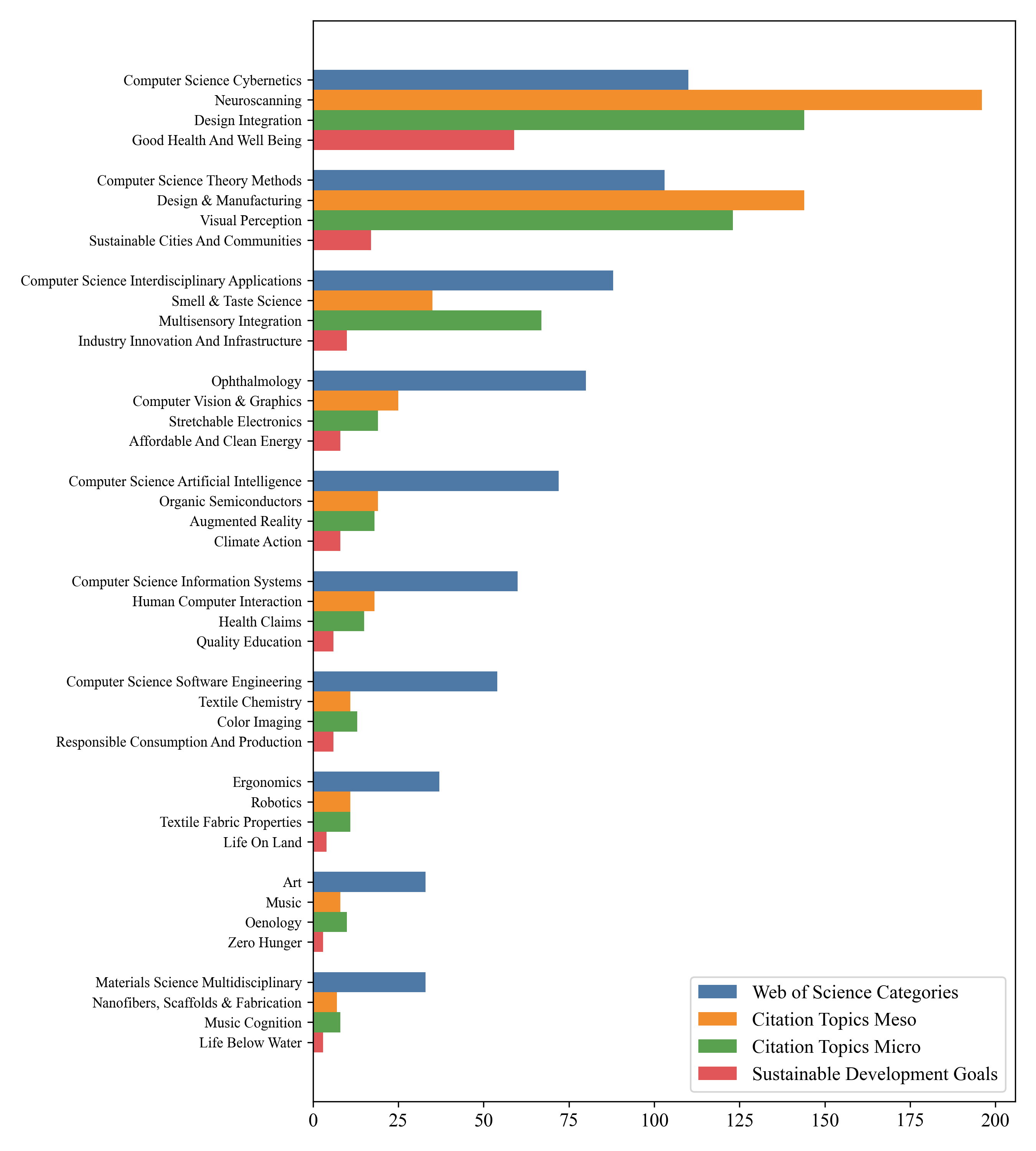}
\caption{Classification Distribution of the Analyzed Literature across Web of Science Categories, Citation Topics Meso, Citation Topics Micro, and Sustainable Development Goals}
\label{figure-4}
\end{figure}

Figure~\ref{figure-4} illustrates the classification distribution of the literature across Web of Science Categories, Citation Topics Meso, Citation Topics Micro, and Sustainable Development Goals. The classification results across different levels similarly reveal the pronounced interdisciplinary nature of material experience research, thereby offering a richer perspective for understanding materials.

\subsection{Country/Region Co‑Authorship Analysis}

Table~\ref{tab:1} lists the top 20 countries/regions ranked by the number of publications and their respective publication counts. The results show that the United States ranks first with 111 publications, followed by China with 98 and Japan with 90. Germany (71 publications), the Netherlands (59 publications), England (53 publications), France (33 publications), Italy (27 publications), Australia (22 publications), and Canada (22 publications) are ranked fourth to tenth, respectively.

\begin{table}[htbp]
\caption{Top 20 Productive Countries/Regions of ``Material Experience'' Publications}
\label{tab:1}
\centering
\begin{tabular*}{\textwidth}{@{\extracolsep{\fill}}lllllllll r@{}}
\toprule
Rank & Count & Centrality & Year & Country & Rank & Count & Centrality & Year & Country \\
\midrule
1  & 111 & 0.13 & 2007 & USA     & 11 & 19 & 0.64 & 2011 & SWE      \\
2  & 98  & 0.25 & 2005 & CHN     & 12 & 17 & 0.29 & 2008 & KOR      \\
3  & 90  & 0.47 & 2005 & JPN     & 13 & 17 & 0.21 & 2008 & DNK      \\
4  & 71  & 0.19 & 2011 & DEU     & 14 & 14 & 0.19 & 2010 & TUR      \\
5  & 59  & 0.20 & 2006 & NLD     & 15 & 10 & 0.10 & 2016 & ESP      \\
6  & 53  & 0.70 & 2006 & ENGLAND & 16 & 7  & 0.00 & 2012 & BEL      \\
7  & 33  & 0.87 & 2006 & FRA     & 17 & 7  & 0.04 & 2014 & AUT      \\
8  & 27  & 0.51 & 2006 & ITA     & 18 & 7  & 0.00 & 2014 & FIN      \\
9  & 22  & 0.00 & 2009 & AUS     & 19 & 7  & 0.07 & 2012 & SCOTLAND \\
10 & 22  & 0.13 & 2012 & CAN     & 20 & 5  & 0.00 & 2013 & NZL      \\
\bottomrule
\end{tabular*}
\end{table}

\begin{figure}[htbp]
\centering
\includegraphics[width=\linewidth]{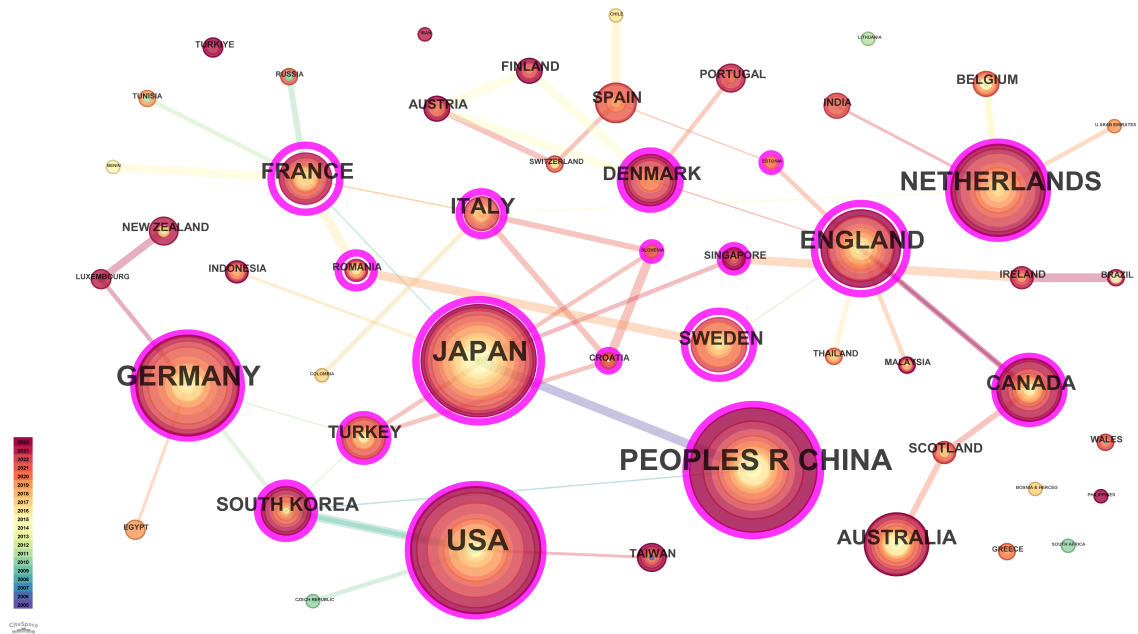}
\caption{Country/Region Co‑Authorship Analysis: Linking Relationships Between Nations}
\label{figure-5}
\end{figure}

Figure~\ref{figure-5} illustrates the network of cooperation between countries/regions. From the perspective of neighboring nodes, the United States maintains close collaborations with South Korea, the Czech Republic, its own domestic institutions, and Taiwan. China primarily maintains strong ties with Japan, South Korea, and its domestic institutions. Japan has established solid collaborative relationships with its domestic institutions, France, China, Indonesia, and Singapore. Germany collaborates with Turkey, Egypt, South Korea, Luxembourg, and its own research institutions. The Netherlands engages in collaborative efforts with Belgium, India, Italy, its domestic institutions, and the United Arab Emirates. From the perspective of centrality, France ranks first with a value of 0.87, followed by England at 0.70, and Romania in third place at 0.66. Ranked fourth to tenth are Sweden (0.64), Italy (0.51), Japan (0.47), South Korea (0.29), China (0.25), Denmark (0.21), and the Netherlands (0.20), respectively. This indicates that countries/regions such as France, England, and Romania have served as intermediaries or hubs across multiple research directions or fields, connecting different research clusters or teams \cite{freemanSetMeasuresCentrality1977}.

Overall, the United States, China, Japan, Germany, and the Netherlands have a large research scale in this field, while France, England, and Romania serve as intermediaries or hubs across multiple research directions or teams, facilitating interdisciplinary exchange and collaboration.

\subsection{Institution Co‑Authorship Analysis}

Table~\ref{tab:2} presents the top ten institutions by number of publications along with their respective countries/regions. The results show that Delft University of Technology in the Netherlands ranks first with 45 publications, followed closely by Justus Liebig University Giessen in Germany with 33 publications. The Centre National de la Recherche Scientifique (CNRS) in France ranks third with 13 publications. In terms of institutional nature, the top ten include not only universities but also a Fortune Global 500 company—Nippon Telegraph and Telephone Corporation (NTT), which ranks fifth with 11 publications. In addition, outside the top ten, several research institutions have also demonstrated active academic output in this field. For example, France’s National Institute for Research in Digital Science and Technology (INRIA), as well as Japan’s National Institute for Physiological Sciences (NIPS) and the National Institutes of Natural Sciences (NINS), have published 7, 6, and 6 papers respectively, ranking joint 13th and joint 17th. These findings suggest that in the context of international collaboration, attention should not be limited to partnerships among universities—enterprises and research institutions also play a vital role. This highlights the ongoing contributions of both industry and research bodies to the cutting-edge field of material experience, driving interdisciplinary integration and technological innovation \cite{soledadduartepobleteEmergingMaterialsFostering2023}.

\begin{table}[htbp]
\caption{Top 20 Productive Institutions of ``Material Experiences'' Publications}
\label{tab:2}
\centering
% 使用 tabular* 使表格宽度等于页面文本宽度 (\textwidth)
% @{\extracolsep{\fill}} 会自动调整列间距，实现“两头顶头”
\begin{tabular*}{\textwidth}{@{\extracolsep{\fill}}lllll@{}}
\toprule
Rank & Count & Year & Institution & Nation \\
\midrule
1  & 45 & 2012 & Delft University of Technology & NLD \\
2  & 33 & 2011 & Justus Liebig University Giessen & DEU \\
3  & 13 & 2006 & Centre National de la Recherche Scientifique (CNRS) & FRA \\
4  & 12 & 2019 & Eindhoven University of Technology & NLD \\
5  & 11 & 2014 & NTT & JPN \\
6  & 10 & 2009 & Massachusetts Institute of Technology (MIT) & USA \\
6  & 10 & 2020 & University of Colorado System & USA \\
8  & 9  & 2014 & American University & USA \\
8  & 9  & 2010 & Cornell University & USA \\
8  & 9  & 2013 & Polytechnic University of Milan & ITA \\
11 & 8  & 2013 & Carnegie Mellon University & USA \\
11 & 8  & 2008 & Tsinghua University & CHN \\
13 & 7  & 2013 & Chiba University & JPN \\
13 & 7  & 2010 & Ihsan Dogramaci Bilkent University & TUR \\
13 & 7  & 2014 & INRIA & FRA \\
13 & 7  & 2019 & University of Tokyo & JPN \\
17 & 6  & 2017 & Lancaster University & ENGLAND \\
17 & 6  & 2011 & National Institute for Physiological Sciences (NIPS) & JPN \\
17 & 6  & 2011 & National Institutes of Natural Sciences (NINS) & JPN \\
17 & 6  & 2014 & Simon Fraser University & CAN \\
\bottomrule
\end{tabular*}
\end{table}

\begin{figure}[htbp]
\centering
\includegraphics[width=\linewidth]{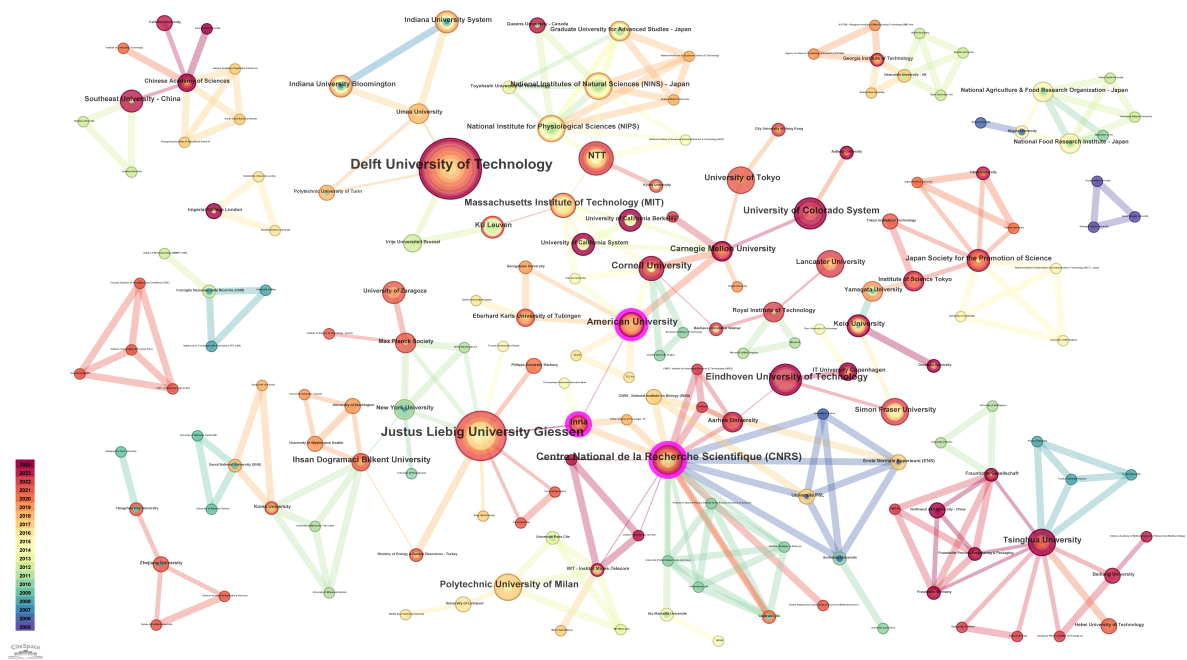}  % 宽度等于正文宽度
\caption{Institution Co‑Authorship Analysis: Linking Relationships Between Institutions}
\label{figure-6}
\end{figure}

Figure~\ref{figure-6} presents the institution co-authorship analysis. According to the analysis of neighboring nodes, Delft University of Technology maintains close collaborations with Umea University, Polytechnic University of Turin, and Vrije Universiteit Brussel. Justus Liebig University Giessen, on the other hand, has established strong collaborative ties with the Max Planck Society, the Ministry of Energy \& Natural Resources of Turkey, New York University, Technical University of Berlin, University of Osnabruck, Edge Hill University, Canon Incorporated, INRIA, Philipps University Marburg, and Canon Australia. We found that Delft University of Technology primarily focuses on collaboration within academia, whereas Justus Liebig University Giessen has a broader scope of cooperation, spanning universities, societies, research institutes, and enterprises.

Based on centrality analysis, we identified American University, INRIA, and CNRS as key nodes with high centrality in the collaboration network. American University maintains close collaborative relationships with Cornell University, Carnegie Mellon University, Georgetown University, Oracle, TCL Inc., Harvard University, INRIA, and Eberhard Karls University of Tübingen. INRIA has established frequent collaborations with American University, Justus Liebig University Giessen, the French National Centre for Scientific Research (Centre National de la Recherche Scientifique, CNRS), its own headquarters, the Communauté Université Grenoble Alpes, and the Istituto Italiano di Tecnologia (IIT). CNRS also demonstrates a broad collaborative network, partnering with the CNRS Institute for Information Sciences and Technologies (INS2I), the Prokhorov General Physics Institute of the Russian Academy of Sciences, the University of Padua, Aix-Marseille Université, École Normale Supérieure (ENS), Autodesk, the French National Institute of Health and Medical Research (Inserm), Centrale Lille, the CNRS National Institute for Biological Sciences (INSB), INRIA, Université de Lille, Université PSL, Institut Mines-Télécom (IMT), Sorbonne Université, Institut Polytechnique de Paris, the CNRS Institute for Humanities and Social Sciences (INSHS), Université Polytechnique Hauts-de-France, Aarhus University, the Russian Academy of Sciences, and IIT. Notably, American University maintains close collaborations with corporate partners such as Oracle and TCL Inc., while CNRS collaborates with Autodesk. These partnerships highlight the unique value that corporate entities bring to this field—they not only manage large-scale user data but also have accumulated extensive expertise in the design and implementation of user experiences. This gives them a significant advantage in user-oriented applications, capabilities that academic institutions typically lack. Therefore, we believe that the active involvement of corporations plays a crucial role in advancing the development of material-related fields  \cite{bontempiSustainableMaterialsTheir2021a}.

In addition, as the only enterprise among the top 20 institutions, Nippon Telegraph and Telephone Corporation (NTT) maintains close collaborative relationships with the National Institute for Physiological Sciences (NIPS), the National Institute of Advanced Industrial Science and Technology (AIST), Kyoto University, the Massachusetts Institute of Technology (MIT), and the National Institutes of Natural Sciences (NINS), most of which are based in Japan.

It is worth noting that Chinese institutions such as Tsinghua University, Zhejiang University, and the Chinese Academy of Sciences primarily collaborate with domestic partners. In the field of material experience, strengthening international research collaboration and industry-academia partnerships can further enhance academic influence.

\subsection{Author Co‑Authorship Analysis}

\begin{figure}[htbp]
\centering
\includegraphics[width=\linewidth]{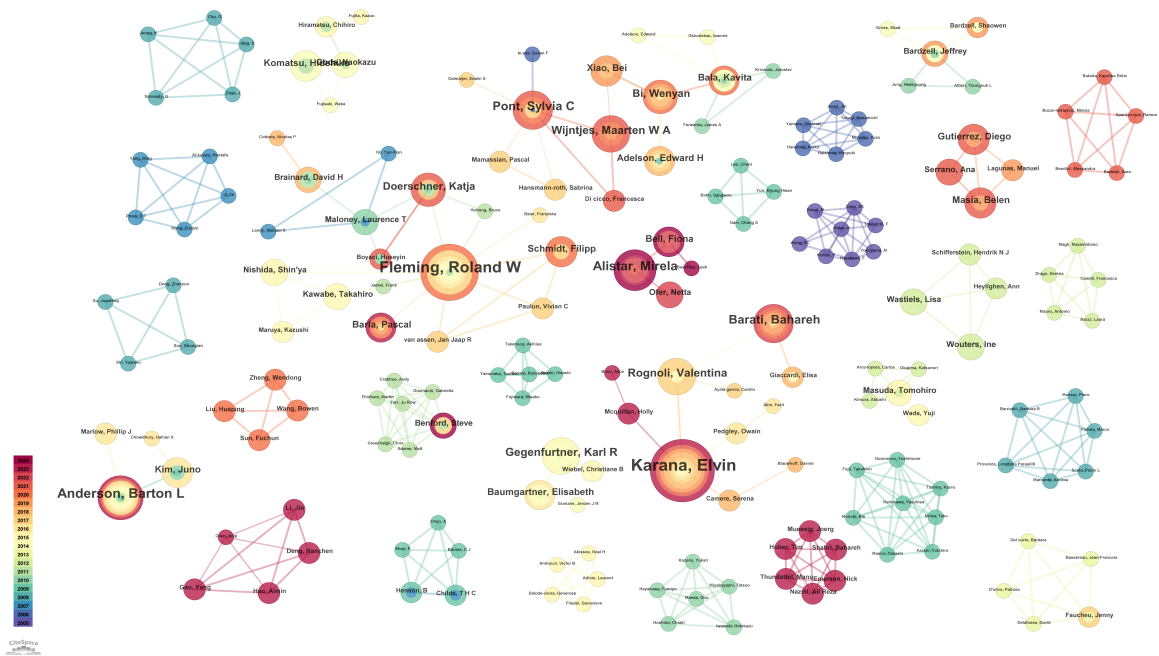}  % 宽度等于正文宽度
\caption{Authors Co‑Authorship Analysis: Linking Relationships Between Authors}
\label{figure-7}
\end{figure}

Table~\ref{tab:3} lists the top 20 most productive authors in the field of “Material Experience” publications, while Figure~\ref{figure-7} presents the Authors Co-Authorship Analysis. 

\begin{table}[htbp]
\caption{Top 20 Productive Authors of ``Material Experience'' Publications}
\label{tab:3}
\centering
\begin{tabular*}{\textwidth}{@{\extracolsep{\fill}}lllll@{}}
\toprule
Rank & Count & Year & Author & Institution (Nation) \\
\midrule
1  & 17 & 2015 & Karana, Elvin & Delft University of Technology (NLD) \\
2  & 14 & 2011 & Fleming, Roland W & Justus Liebig University Giessen (DEU) \\
3  & 9  & 2009 & Anderson, Barton L & The University of Sydney (AUS) \\
4  & 8  & 2020 & Alistar, Mirela & University of Colorado System (USA) \\
5  & 7  & 2006 & Pont, Sylvia C & Delft University of Technology (NLD) \\
5  & 7  & 2015 & Barati, Bahareh & Eindhoven University of Technology (NLD) \\
7  & 6  & 2019 & Wijntjes, Maarten W A & Delft University of Technology (NLD) \\
7  & 6  & 2015 & Rognoli, Valentina & Polytechnic of Milan (ITA) \\
7  & 6  & 2013 & Gegenfurtner, Karl R & Justus Liebig University Giessen (DEU) \\
10 & 5  & 2015 & Balas, Benjamin & North Dakota State University (USA) \\
10 & 5  & 2016 & Bi, Wenyan & Yale University (USA) \\
10 & 5  & 2010 & Doerschner, Katja & Justus Liebig University Giessen (DEU) \\
13 & 4  & 2017 & Barla, Pascal & Universite de Bordeaux (FRA) \\
13 & 4  & 2018 & Xiao, Bei & American University (USA) \\
13 & 4  & 2016 & Faul, Franz & University of Kiel (DEU) \\
13 & 4  & 2013 & Baumgartner, Elisabeth & Justus Liebig University Giessen (DEU) \\
13 & 4  & 2021 & Bell, Fiona & University of Colorado System (USA) \\
13 & 4  & 2019 & Masia, Belen & University of Zaragoza (ESP) \\
13 & 4  & 2009 & Adelson, Edward H & Massachusetts Institute of Technology (USA) \\
13 & 4  & 2010 & Bala, Kavita & Cornell University (USA) \\
\bottomrule
\end{tabular*}
\end{table}

Among them, Elvin Karana from Delft University of Technology ranks first with 17 papers. Her research explores the integration of design, biotechnology, and materials science, aiming to create innovative biomaterials that harmoniously fit into daily human life while accommodating the diverse cycles, scales, and timing of natural ecosystems. She is the founder of the Delft Bio Design Lab and, together with Professor Valentina Rognoli at Politecnico di Milano, co-directs the Material Experiences Lab \cite{tudelft-karanae.ProfDrKarana}. Her 2015 paper Material Driven Design (MDD): A Method to Design for Material Experiences, published in the International Journal of Design, has exerted extensive influence \cite{karanaMaterialDrivenDesign2015}. She advocates viewing materials as a medium — one that can convey ideas, beliefs, and methods, prompting people to form specific patterns of thinking, feeling, and acting, while also enabling and enhancing the functionality and practicality of products \cite{karanaMaterialDrivenDesign2015, karanaMaterialsExperience2015a, cameraExperientialCharacterizationMaterials2018, giaccardiFoundationsMaterialsExperience2015a, karanaWhenMaterialGrows2018}. Roland W. Fleming from Justus Liebig University Giessen ranks second with 14 publications. His research focuses on the visual perception of objects and materials, with particular attention to how the physical properties of surfaces, materials, and objects in the surrounding environment can be inferred through vision. It aims to reveal how the brain estimates the three-dimensional shapes and physical attributes of objects and materials, as well as the impact of this process on perception, imagination, and action \cite{giessenuniversity-rolandflemingProfRolandFleming, flemingMaterialPerception2017, flemingSpecularReflectionsPerception2004}. To this end, he comprehensively employs a variety of methods such as computer graphics, image analysis technology, neural modeling, behavioral experiments, and brain imaging technology \cite{giessenuniversity-rolandflemingProfRolandFleming}. His paper Visual perception of materials and their properties, published in Vision Research, embodies his academic ideas in a concentrated manner and has far-reaching influence\cite{flemingVisualPerceptionMaterials2014}. Barton L. Anderson from the University of Sydney ranks third with nine publications. His primary research focuses on perceptual understanding, with a particular emphasis on visual appearance. His work covers perceptual organization, segmentation, grouping, and the recovery of surface properties such as color, material composition, and shape \cite{theuniversityofsydneyProfessorBartAnderson}. His representative works include “Image statistics do not explain the perception of gloss and lightness”, “Image statistics and the perception of surface gloss and lightness” and “Generative constraints on image cues for perceived gloss” \cite{andersonImageStatisticsNot2009, kimImageStatisticsPerception2010, marlowGenerativeConstraintsImage2013}. Ranking fourth is Mirela Alistar from the University of Colorado System, with a total of eight publications. She is an active advocate of the DIY Bio (Do-It-Yourself Biology) movement and has led and co-founded several community wet labs. Mirela Alistar is also the director of the Living Matter Lab at the ATLAS Institute. Within this context, she has organized various interactive performances, art installations, and open workshops aimed at engaging the public in direct interaction with living materials such as bacteria, viruses, and fungi \cite{universityofcoloradoboulderMirelaAlistar}. Her representative works include “Living Media Interfaces: A Multi-Perspective Analysis of Biological Materials for Interaction” and “Designing Direct Interactions with Bioluminescent Algae” \cite{merrittLivingMediaInterfaces2020, oferDesigningDirectInteractions2021}.

Sylvia C. Pont from Delft University of Technology and Bahareh Barati from Eindhoven University of Technology are tied for fifth place, each with seven publications. Sylvia C. Pont is the head of the Perceptual Intelligence Lab ($\pi$-lab), focusing on multisensory experiences and the perception and design of products, services, and systems. She is known for her interdisciplinary approach to addressing real-world perceptual challenges. Her research primarily centers on perception-based lighting design, ecological optics of light, and the interactions between light, materials, shapes, and spatial configurations \cite{tudelft-s.c.sylviapontProfdrSCSylvia, koenderinkVisualLightField2007, pontLightTransdisciplinaryScience2019}. Bahareh Barati's research focuses on the creative processes associated with emerging materials, including smart and biological materials, and delves into the unique role designers play in multidisciplinary material development. She has developed digital tools to support the prototyping of novel composite materials, fostering interdisciplinary collaboration and promoting innovative material exploration \cite{eindhovenuniversityoftechnologyBaharehBarati, baratiPrototypingMaterialsExperience2019}. In addition, she is dedicated to expanding the role of digital technologies—such as digital twins and artificial intelligence—in helping us understand and explore the everyday practices of using, through and around (emerging) living materials such as bacteria, fungi, and algae \cite{eindhovenuniversityoftechnologyBaharehBarati, zhouHabitabilitiesLivingArtefacts2022}. Notably, she co-developed the concept of Material Driven Design (MDD) with Elvin Karana and has carried out extensive practical and design work in this area \cite{karanaMaterialDrivenDesign2015}.

Tied for seventh place are three scholars, each with six publications: Maarten W. A. Wijntjes from Delft University of Technology, Valentina Rognoli from the Polytechnic University of Milan, and Karl R. Gegenfurtner from Justus Liebig University Giessen. Maarten W. A. Wijntjes primarily focus on visual perception and communication design, particularly on the modeling of visual perception, investigating perceptual phenomena that can be described through physical or mathematical models — such as light, space, shape, color, and material properties \cite{tudelft-maartenw.a.wijntjesMaartenWijntjes}. His approaches these issues from a design perspective, exploring how visual communication and the depiction of materials convey perceptual qualities \cite{wijntjesIllusoryGlossLambertian2010, wijntjesVisualCommunicationHow2019, vanzuijlenPainterlyDepictionMaterial2020}. Valentina Rognoli’s research spans a broader range of topics in materials and design, including bio-based and biofabricated materials, materials derived from waste and food waste, materials for interaction and the Internet of Things (ICS materials), speculative materials, tinkering with materials, the Material Driven Design (MDD) method, CMF design, emerging material experiences, and material education in the field of design \cite{materialsexperiencelabPROFDRVALENTINA}. She is the founder and lead researcher of the Materials Design for Transition research group, established in 2023, which focuses on tinkering and biotinkering practices with emerging materials \cite{politecnicodimilanoValentinaRognoli, rognoliDIYMaterials2015, rognoliMaterialsBiographyTool2022, ayala-garciaNewAestheticDIYMaterials2017, parisiMaterialTinkeringInspirational2017}. She is also a key contributor to the development of the MDD concept, co-proposed with Elvin Karana \cite{karanaMaterialDrivenDesign2015}. Karl R. Gegenfurtner's work is rooted in information processing within the visual system. Specifically, his research focuses on the relationship between low-level sensory processes, higher-level visual cognition, and sensorimotor integration \cite{giessenuniversity-karlr.gegenfurtnerProfKarlGegenfurtner, gegenfurtnerColorVision2003, gegenfurtnerCorticalMechanismsColour2003}.

Three scholars are tied for tenth place, each with five publications: Benjamin Balas from North Dakota State University, Wenyan Bi from Yale University, and Katja Doerschner from Justus Liebig University Giessen.

We found that continental European countries/regions, especially the Netherlands, Germany, and Italy, demonstrate a strong leading advantage in this field; meanwhile, the United States, France, and Japan also maintain a certain level of competitiveness. On the other hand, although China has made significant progress and rapid development in this area, there remains a gap in the contribution of core theories and the production of high-quality publications. In addition, scholars such as Elvin Karana, Valentina Rognoli, and Bahareh Barati not only have a background in design but also actively integrate knowledge from materials science, biological sciences, and artificial intelligence. This reflects a shift in material experience research from a traditional design perspective toward a cross-disciplinary model that blends design, technology, and the humanities. Researchers like Sylvia C. Pont, Roland W. Fleming, and Barton L. Anderson focus on visual perception and the interaction between light and materials, demonstrating that traditional perceptual science is providing objective and quantifiable theoretical support for material experience research. The Material Driven Design (MDD) method, proposed by Elvin Karana and others, has become a key theoretical framework in this field and is being continuously expanded and applied by an increasing number of scholars. Meanwhile, researchers such as Bahareh Barati and Mirela Alistar actively introduce living materials such as bacteria, algae, and fungi into their work, combining biological interaction, public engagement, and DIY biology practices. This indicates that material experience research is increasingly intertwined with issues such as sustainable design, ecological ethics, and public co-creation, fostering a shift in our understanding of materials—from physical substance to living matter.

\subsection{Co‑Occurrence Analysis for the Research Frontier and Trends}

\begin{figure}[htbp]
\centering
\includegraphics[width=\linewidth]{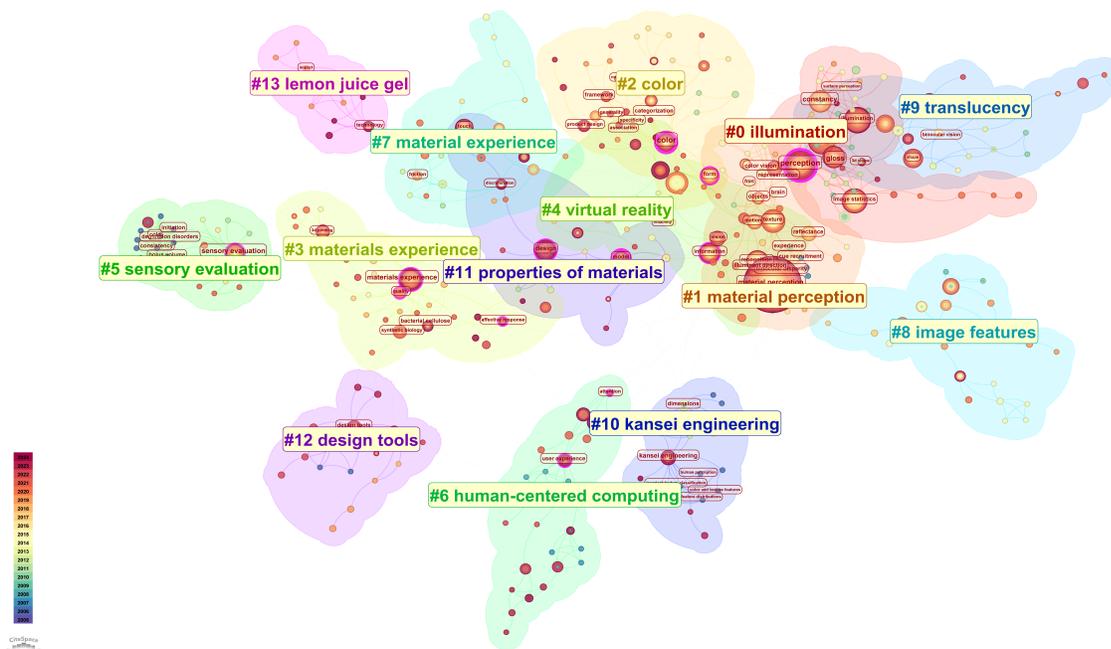}
\caption{The network structure of 14 keyword clusters, where nodes represent keywords and links indicate co-occurrence relationships related to “Material Experiences”}
\label{figure-8}
\end{figure}

To gain a deeper understanding of the major developmental themes and the knowledge structure within the field of materials experience research, this study conducted a keyword clustering analysis. As shown in Figure~\ref{figure-8}, a total of 14 valid clusters were identified, with a modularity (Q) value of 0.83 and an average silhouette score of 0.93, indicating a clear clustering structure with high validity and consistency. These clusters include: \#0 Illumination, \#1 Material Perception, \#2 Color, \#3 Materials Experience, \#4 Virtual Reality, \#5 Sensory Evaluation, \#6 Human-Centered Computing, \#7 Material Experience, \#8 Image Features, \#9 Translucency, \#10 Kansei Engineering, \#11 Properties of Materials, \#12 Design Tools, and \#13 Lemon Juice Gel.

\begin{figure}[htbp]
\centering
\includegraphics[width=\linewidth]{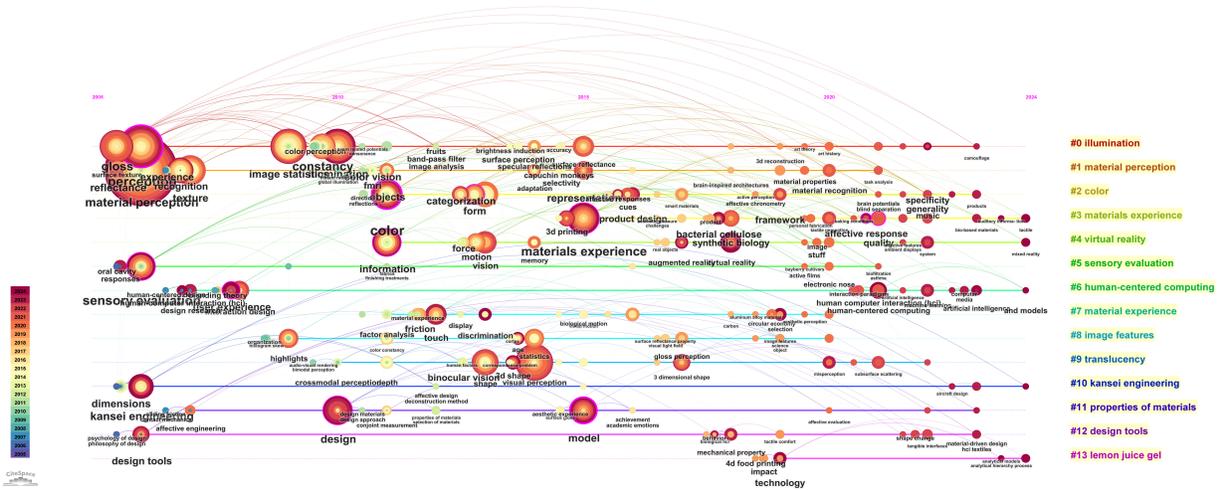}
\caption{The evolution of keywords over time (Timeline View), where lines of different colors represent keywords from different clusters, and the X-axis represents the time period (2005 - 2024)}
\label{figure-9}
\end{figure}

Figure~\ref{figure-9} illustrates the temporal evolution of keyword clusters. Early research on material experience primarily focused on the visual perception and physical properties of materials, reflecting a material-centered research perspective. As the field developed, research themes gradually diversified, encompassing multimodal perception, emotional experience, user interaction, and design applications. Meanwhile, the introduction of computational methods, virtual reality, and intelligent design tools has propelled material experience research toward more quantitative and technologically driven directions. This evolution indicates that material experience research is gradually shifting from a material-centered perspective to a user-centered one, while simultaneously exhibiting a trend of coexistence of multiple perspectives.

Table~\ref{tab:4} summarizes the results of the keyword clustering analysis, presenting each cluster’s thematic labels, number of keywords (size), cluster quality metric (silhouette score), representative year, and main related keywords. 

\begin{sidewaystable}[htbp]
\caption{The Results of the Keyword Clustering Analysis}
\label{tab:4}
\centering
\begin{tabularx}{\textheight}{@{}l c c c X@{}}
\toprule
Cluster ID & Size & Silhouette & Year & Label (LLR) \\
\midrule
\#0 illumination & 42 & 0.897 & 2013 & illumination (8.13, 0.005); color perception (7.01, 0.01); art history (7.01, 0.01); material attributes (7.01, 0.01); lightness (6.41, 0.05) \\ \addlinespace[3pt]
\#1 material perception & 34 & 0.894 & 2012 & material perception (10.25, 0.005); texture (9.16, 0.005); material properties (8.28, 0.005); fmri (6.81, 0.01); correlation (6.4, 0.05) \\ \addlinespace[3pt]
\#2 color & 34 & 0.946 & 2017 & color (12.44, 0.001); affective interfaces (4.9, 0.05); evaluation (4.9, 0.05); music (4.9, 0.05); odor activity value (4.9, 0.05) \\ \addlinespace[3pt]
\#3 materials experience & 32 & 0.904 & 2019 & materials experience (35.39, 1.0E-4); material perception (16.79, 1.0E-4); affective response (15.41, 1.0E-4); 3d printing (15.41, 1.0E-4); biodesign (10.25, 0.005) \\ \addlinespace[3pt]
\#4 virtual reality & 30 & 0.881 & 2017 & virtual reality (12.75, 0.001); augmented reality (9.71, 0.005); haptic i/o (4.84, 0.05); keyword search (4.84, 0.05); situatedness (4.84, 0.05) \\ \addlinespace[3pt]
\#5 sensory evaluation & 26 & 0.999 & 2011 & sensory evaluation (29.19, 1.0E-4); electronic nose (20.56, 1.0E-4); material perception (7.1, 0.01); trained panel (6.8, 0.01); textile industry (6.8, 0.01) \\ \addlinespace[3pt]
\#6 human-centered computing & 26 & 0.978 & 2015 & human-centered computing (44.08, 1.0E-4); human computer interaction (hci) (37.66, 1.0E-4); interaction paradigms (18.66, 1.0E-4); user experience (14.25, 0.001); machine learning (12.4, 0.001) \\ \addlinespace[3pt]
\#7 material experience & 23 & 0.864 & 2014 & material experience (9.42, 0.005); touch (6.66, 0.01); perceived hardness of solid objects (6.56, 0.05); pleasantness (6.56, 0.05); sustainable luxury eco-innovation (6.56, 0.05) \\ \addlinespace[3pt]
\#8 image features & 23 & 0.929 & 2015 & image features (13.16, 0.001); food choice (13.16, 0.001); gloss perception (6.66, 0.01); nutrition counseling (6.56, 0.05); focus group research (6.56, 0.05) \\ \addlinespace[3pt]
\#9 translucency & 22 & 0.927 & 2014 & translucency (19.47, 1.0E-4); 3d shape (14.9, 0.001); experimentation (14.77, 0.001); binocular vision (10.45, 0.005); human factors (9.82, 0.005) \\ \addlinespace[3pt]
\#10 kansei engineering & 21 & 0.936 & 2011 & kansei engineering (26.92, 1.0E-4); eye-tracking technology (7.33, 0.01); yarn design (7.33, 0.01); visualization network (7.33, 0.01); cmf (7.33, 0.01) \\ \addlinespace[3pt]
\#11 properties of materials & 21 & 0.983 & 2013 & properties of materials (11.39, 0.001); affective engineering (11.39, 0.001); selection of materials (11.39, 0.001); material perception (6.76, 0.01); blind and low vision music learning (5.68, 0.05) \\ \addlinespace[3pt]
\#12 design tool & 17 & 0.969 & 2016 & design tools (30.05, 1.0E-4); shape-changing interfaces (7.4, 0.01); prototyping (7.4, 0.01); soft robotics (7.4, 0.01); fabrication (7.4, 0.01) \\ \addlinespace[3pt]
\#13 lemon juice gel & 11 & 0.992 & 2020 & lemon juice gel (9.4, 0.005); notes (9.4, 0.005); health food product (9.4, 0.005); smart material (9.4, 0.005); chinese sweet orange (9.4, 0.005) \\
\bottomrule
\end{tabularx}
\par\smallskip
\footnotesize
Note: LLR = Log-Likelihood Ratio; for each keyword, the first number in parentheses represents its LLR value, indicating its representativeness within the cluster, and the second number is the p-value, indicating the statistical significance of the keyword's association with the cluster.
\end{sidewaystable}

Specifically, early clusters such as the Kansei Engineering cluster, emphasized the integration of technology and design. Following this, the Sensory Evaluation cluster and the Material Perception cluster marked the shift in material experience research from a singular sensory perspective to a multisensory approach, beginning to incorporate more scientific measurement techniques. As research progressed, the Illumination cluster and the Properties of Materials cluster reflected a focus on emotional experiences and the needs of special groups. The Translucency cluster demonstrated the deep integration of technologies, pushing material experience research toward more practical directions. The Material Experience cluster highlighted a user-centered approach in material experience research, emphasizing the close connection between materials and user emotions and perceptions. Overall, these stages illustrate the evolution of material experience research from a focus on materials themselves to a more user-centered research orientation, reflecting the gradual application of scientific methods and driving research toward a more integrated, diverse, and human-centered direction.

In recent years, with the integration of technologies such as Human-Centered Computing, Design Tools, and Virtual Reality, material experience research has shown a clear trend of interdisciplinary integration, driving the development of smart interaction and digital design. At the same time, the Image Features cluster and the Color cluster demonstrate in-depth exploration within traditional fields. Additionally, the Materials Experience cluster and the Lemon Juice Gel cluster represent emerging interdisciplinary explorations between biodesign, smart materials, and the food-health field. In conclusion, material experience is shifting from traditional “materiality” to materials endowed with “vitality” and “intelligence”.

Overall, these 14 clusters not only cover the core topics of material experience but also demonstrate a multilayered, cross-domain integration—from perceptual science to new material technologies, and further to socio-cultural and humanistic concerns—validating the field's ongoing evolution as an interdisciplinary domain connecting design, technology, and the humanities, and providing a clear direction and solid theoretical foundation for future development.

\begin{figure}[t]
\centering
\includegraphics[scale=0.8]{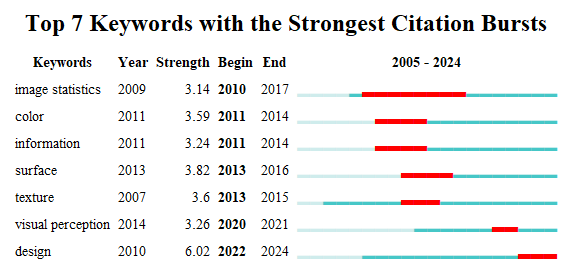}
\caption{Keywords with the Strongest Citation Bursts in Material Experience Research}
\label{figure-10}
\end{figure}

Figure~\ref{figure-10} presents the results of the keyword citation burst analysis, highlighting keywords that experienced a significant increase in frequency in material experience research between 2005 and 2024. Notably, terms such as “image statistics”, “color”, “information”, “surface”, “texture”, “visual perception” and “design” exhibited strong bursts in specific years, indicating that these topics became focal points of scholarly attention during those periods. The recent bursts of keywords like “information” and “design” suggest a shift in research focus toward experiential design and digital dimensions, reflecting the growing trend of integrating design methods with information technology in material experience.

The evolution of research hotspots reveals that early studies in material experience primarily focused on the fundamental perceptual characteristics of materials, such as “image statistics” (2010–2017), “color” (2011–2014), “surface” (2013–2016), and “texture” (2013–2015). These keywords were concentrated on the visual and tactile dimensions, as well as foundational research methods, reflecting the academic community's strong interest at the time in the appearance-related properties of materials and their perceptual mechanisms. Since 2020, the research focus has gradually shifted from material appearance properties to deeper cognitive processes. For example, the emergence of the keyword “visual perception” (2020–2021) marks a deeper exploration of users' perceptual processes, while “design” (2022–2024), with a peak burst strength of 6.02, highlights its transition from a supporting method to a central driving force in material experience research. This trend not only reflects the shift in research paradigms from singular sensory approaches to multimodal integration but also emphasizes the bridging role of design in connecting sensory experience, emotional resonance, and practical value. Design is no longer merely a tool for expressing perceptual outcomes; it has become a key mediator that coordinates sensory input, user cognition, and contextual needs. Overall, this transformation marks a significant shift in material experience research from “perceiving materials” to “designing experiences,” with design emerging as a core force driving continuous innovation in the field.

\section{Discussion}
\subsection{The Disappearance of Material Boundaries}
Although “breaking the boundaries of materiality” has become one of the future trends in the field of material experience, as designers, we must examine the contradictions and risks behind the dissolution of these boundaries. The experience design of biomaterials blurs the line between “material and living organism,” which could lead to ethical debates. For example, the “right to life” of living materials, issues of social acceptance, and the need for the development of legal and moral frameworks. The experience of “immaterial materials” may gradually dull users' perception of real physical materials, potentially leading to sensory detachment and a lack of emotional resonance. In the face of these challenges, we believe that designers should approach material experience from the perspective of fundamental science, rather than merely focusing on the design aspect.

\subsection{The ``non-quantifiability'' and subjectivity of experience}
Material experience research is gradually moving beyond its traditional reliance on materials science, with researchers increasingly focusing on the computability of experiences, aiming to translate abstract emotional perceptions into explicit mathematical models. However, compared to natural sciences, the conclusions of these studies remain difficult to reproduce or falsify. The root cause lies in the underestimation of the “non-quantifiability” of emotions—emotional experiences are fluid and heavily influenced by individual differences and other factors. This also leads to issues such as biases in data sources, which continue to affect the generalizability of research findings. We believe that with the development of information technology, artificial intelligence, and big data, adopting an interdisciplinary research perspective can effectively enhance the stability and reliability of conclusions.

\subsection{Material innovation driven by design or by technology}
Especially in the context of today's artificial intelligence era, whether material experience is driven by technology or design has become a key issue. At the same time, in the context of interdisciplinary integration, the subjectivity of design also faces challenges. However, the author believes that by adhering to a human-centered design philosophy and avoiding the blind labeling of materials, we should seek a balance between “technological possibilities” and “humanistic necessities”. This reflects the duality of materials, balancing their practicality with the pleasure they provide to humans. This is the core role of design in material experience.

\section{Conclusions}

This study identifies several key developmental trends in the field of material experience, as outlined below:
\begin{itemize}
    \item The scope of research has significantly expanded. The focus of research has gradually shifted towards emerging fields such as intangible materials, virtual materials, smart materials, and biomaterials. In particular, the integration of physical matter into non-material biological systems and the development of programmable materials has shown tremendous potential, offering a new direction for future material innovation. 
    \item There has been a significant shift in research methods. The focus of research has gradually shifted from subjective descriptions of experience to experimental approaches based on objective data measurement, incorporating techniques such as mathematical modeling, machine learning, and deep learning. This marks a continuous improvement in scientific rigor and quantitative analysis within the field, providing a more solid theoretical foundation for the in-depth study of material experience.
    \item Multimodal interaction and multisensory integration. This trend not only enriches the perceptual layers of material experience but has also been extensively explored and applied in virtual environments and smart material applications.
    \item Enhanced role of design and interdisciplinary integration. This trend indicates that design is gradually becoming a central mediator, coordinating sensory input, user cognition, and environmental demands, while research is increasingly focusing on emotional, cultural, ecological, and sustainability dimensions.
\end{itemize}

Overall, these trends indicate that material experience research is undergoing a comprehensive transformation. The bibliometric analysis in this study provides a systematic overview for understanding the evolution of the field and emerging hotspots. However, this overview mainly reflects the macro-level perspective, and the finer details still need to be grasped and interpreted by researchers in practice.

\section{Statements and Declarations}
    The authors have no competing interests to declare that are relevant to the content of this article.

\bibliographystyle{unsrt}
\bibliography{references}

\end{document}